\documentclass[aps,preprint]{revtex4}%
\usepackage{amsfonts}
\usepackage{amsmath}
\usepackage{amssymb}
\usepackage{graphicx}%
\begin{document}
\preprint{ }
\title[Renormalization of EE from topological boundary terms.]{Renormalization of Entanglement Entropy from topological terms}
\author{Giorgos Anastasiou}
\affiliation{Departamento de Ciencias F\'isicas, Universidad Andr\'es Bello, Sazi\'e 2212,
Piso 7, Santiago, Chile}
\author{Ignacio J. Araya}
\affiliation{Departamento de Ciencias F\'isicas, Universidad Andr\'es Bello, Sazi\'e 2212,
Piso 7, Santiago, Chile}
\author{Rodrigo Olea}
\affiliation{Departamento de Ciencias F\'isicas, Universidad Andr\'es Bello, Sazi\'e 2212,
Piso 7, Santiago, Chile}
\pacs{PACS number}

\begin{abstract}
We propose a renormalization scheme for Entanglement Entropy of 3D CFTs with a
4D asymptotically AdS gravity dual in the context of the gauge/gravity
correspondence. The procedure consists in adding the Chern form as a boundary
term to the area functional of the Ryu-Takayanagi minimal surface. We provide
an explicit prescription for the renormalized Entanglement Entropy, which is
derived via the replica trick. This is achieved by considering a Euclidean
gravitational action renormalized by the addition of the Chern form at the
spacetime boundary, evaluated in the conically-singular replica manifold. We
show that the addition of this boundary term cancels the divergent part of the
Entanglement Entropy, recovering the results obtained by Taylor and Woodhead.
We comment on how this prescription for renormalizing the Entanglement Entopy
is in line with the general program of topological renormalization in
asymptotically AdS gravity.

\end{abstract}
\volumeyear{ }
\startpage{1}
\endpage{2}
\maketitle
\tableofcontents

\section{Introduction}

In the context of the AdS/CFT correspondence \cite{AdS/CFT1}-\cite{AdS/CFT3},
the Entanglement Entropy (EE) of an entangling region $A$ in a CFT with an
asymptotically AdS (AAdS) Einstein gravity dual, can be computed as the volume
of a codimension-2 minimal surface. In particular, this is achieved by
calculating the volume of the minimal surface $\Sigma$ in the bulk whose
boundary $\partial\Sigma$ is conformal to the entangling surface $\partial A$,
which bounds $A$ at the conformal boundary $C$. This proposal is referred to
as the Ryu-Takayanagi (RT) prescription \cite{RT1}\cite{TakayanagiReview}. In
order to illustrate the different submanifolds involved in this construction,
and the geometric relations between them, we include a schematic diagram in
FIG. 1.

\begin{figure}[h]
\begin{center}
\includegraphics[
height=6.0in,
width=6.0in
]{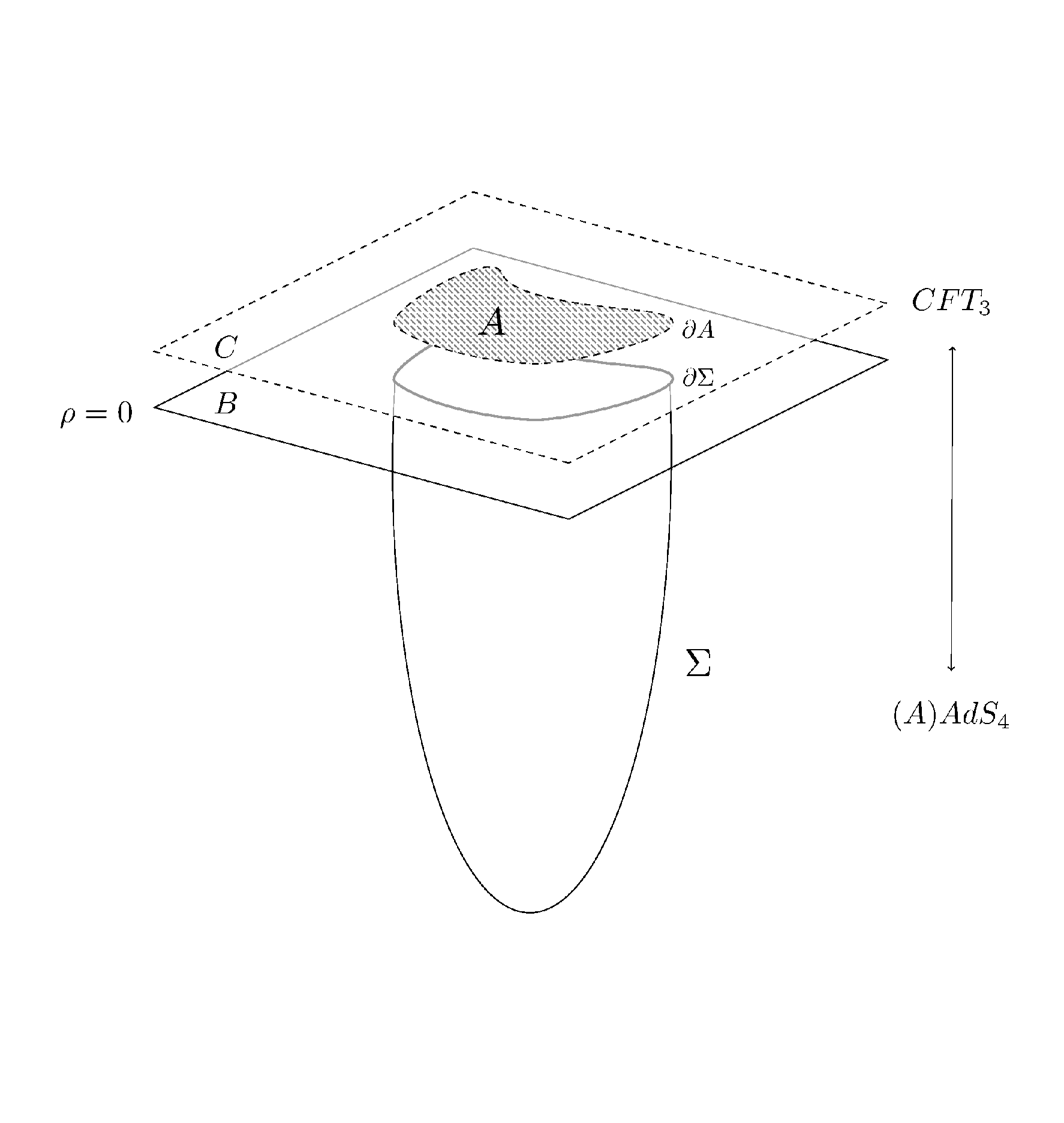}
\end{center}
\caption{In this diagram, we show all the submanifolds involved in the RT
construction. On the field theory side, $C$ is the conformal boundary where
the CFT is defined, $A$ is the entangling region and $\partial A$ is the
entangling surface. On the gravity side, $B$ is the boundary of spacetime,
$\Sigma$ is the minimal surface in the bulk and $\partial\Sigma$ is its border
at the spacetime boundary. Both sides are related such that $C$ is conformal
to $B$ and $\partial A$ is conformal to $\partial\Sigma$.}%
\end{figure}

This definition for the EE is formally divergent, due to the presence of an
infinite conformal factor at the AdS boundary $B$, what is manifest in the
Fefferman-Graham (FG) form of the metric \cite{FeffermanGraham}\cite{Imbimbo}.
As it was shown by Taylor and Woodhead \cite{Marika}, it is possible to
renormalize the EE by adding counterterms constructed through the Replica
Trick \cite{EECovariant}-\cite{Marika} from the standard Holographic
Renormalization procedure \cite{UsualCounterJohnson}-\cite{SkenderisAndPapa2}.
This is done by evaluating the usual counterterms for Einstein gravity at the
conically singular spacetime boundary, which is conformal to the manifold of
the Replica CFT.

Here, we propose an alternative regularization prescription that has the
advantage of giving the countertem for the EE as a single boundary term, which
can be written in closed form for CFTs of arbitrary (odd) dimensions that have
an (even-dimensional) AAdS E-H gravity dual. This boundary term corresponds to
the Chern form evaluated at the boundary of the RT minimal surface, which is
conformal to the entangling surface that bounds the entangling region in the
CFT. In particular, we propose that the renormalized EE of a 3D CFT with a 4D
AAdS E-H gravity dual is given by%

\begin{equation}
S_{EE}^{ren}=\frac{Vol\left(  \Sigma\right)  }{4G}+\frac{\ell^{2}}{8G}%
\int\limits_{\partial\Sigma}B_{1}, \label{S_EE^ren}%
\end{equation}
where $\Sigma$ is the codimension-2 RT minimal surface, $\partial\Sigma$ is
its border at the spacetime boundary $B$, $\ell$ is the AdS radius. In
addition, $B_{1}$ is the first Chern form evaluated at the border of the RT
minimal surface, whose detailed form is given in eq.(\ref{4D_EE_terms}).
Therefore, we show that the EE counterterm ($S_{EE}^{ct}$) is given by the
$B_{1}$ term, which depends on the induced metric $\widetilde{\gamma}$ of
$\partial\Sigma$, and on its extrinsic curvature with respect to the radial
foliation along the holographic radial coordinate $\rho$, which is the
parameter of the FG expansion. It is apparent then, that the Chern form is
written in terms of both intrinsic and extrinsic quantities of $\partial
\Sigma$. The particular features of $S_{EE}^{ct}$ are explained in section
\ref{Renorm}.

In order to obtain the EE boundary counterterm, we consider the Replica Trick,
where the conically singular replica manifold is constructed as described in
\cite{RenyiXiDong}. We also consider the gravitational Euclidean action in the
AAdS bulk. The EE is then expressed in terms of a derivative of said action
evaluated on the replica manifold, with respect to the conical angular
parameter. Therefore, if the action is itself renormalized, the EE computed in
this manner will be renormalized as well. In order to renormalize the bulk
gravitational action, we consider the \textit{Kounterterms} proposal
\cite{K1Even}-\cite{KounterComparison2}, instead of the standard Holographic
Renormalization prescription \cite{UsualCounterJohnson}%
-\cite{SkenderisAndPapa2}. This choice is made because of the fact that in the
Kounterterms scheme, the counterterm that renormalizes the on-shell
gravitational action can be written in closed form, as a single boundary term
with topological origin. As a matter of fact, this prescription is known for
arbitrary dimensions and also for any gravity theory of Lovelock type. For the
evaluation of the renormalized action on the replica manifold, we consider a
generalization of the Euler theorem to conically singular manifolds in 4D,
derived by using distributional geometry \cite{Solodukhin1}%
\cite{SolodukhinNew}. Thus, the counterterm of the action splits into a
regular part at the spacetime boundary and another at $\partial\Sigma$. The
latter results in a contribution proportional to the angular parameter which
gives the $B_{1}$ piece of eq.(\ref{S_EE^ren}). Upon taking the derivative of
the action with respect to the conically singular parameter, we obtain
$S_{EE}^{ren}$ as shown in eq.(\ref{S_EE^ren}), where the bulk part of the
action gives the usual RT term. We emphasize that the form of $S_{EE}^{ren}$
obtained in the AdS$_{4}$/CFT$_{3}$ case, and shown in eq.(\ref{S_EE^ren}), is
equivalent to the known result of%

\begin{equation}
S_{EE}^{ren}=\frac{Vol\left(  \Sigma\right)  }{4G}-\frac{\ell}{4G}%
\int\limits_{\partial\Sigma}dx\sqrt{\widetilde{\gamma}},
\label{S_ren_usual_4D}%
\end{equation}
given in \cite{Marika}, as explained in section \ref{FinitenessProof}.

This paper is organized as follows: In section \ref{Setup}, we explain the
setup used for obtaining the renormalized EE. We give a general overview of
the definition of EE and of the Replica Trick, applied to the AdS/CFT context.
We then explain the generalization of the Euler theorem for conically singular
manifolds, and in particular, for the case without a $U\left(  1\right)  $
isometry in 4D. Then, we introduce the renormalized Euclidean gravitational
action obtained by the Kounterterms procedure. In section \ref{Renorm}, we use
the elements described in the setup to obtain the $S_{EE}^{ct}$ in the
AdS$_{4}$/CFT$_{3}$ case. We also expand the obtained boundary counterterm
considering the explicit covariant embedding of the $\Sigma$ minimal surface
on the bulk. When taking the FG expansion of its induced metric we show that
$S_{EE}^{ct}$ can be re-written in the standard way of
eq.(\ref{S_ren_usual_4D}). We explicitly check the finiteness of $S_{EE}%
^{ren}$, and we verify that the standard computation of the renormalized EE of
a disc-like entangling region in CFT$_{3}$ is correctly recovered. We also
give a new interpretation of $S_{EE}^{ren}$ in terms of the topological and
geometrical properties of the minimal surface $\Sigma$ as an AAdS submanifold
(see eq.(\ref{S_EE^Topol})). Finally, in section \ref{Outlook}, we give a
general outlook of the method and comment on possible generalizations thereof.

\section{The setup: Replica trick and renormalized Euclidean action in the
conically singular manifold\label{Setup}}

We proceed to explain the different elements of the setup considered in order
to obtain the renormalized entanglement entropy $S_{EE}^{ren}$. We start by
giving a brief overview of EE in the AdS/CFT context, discussing how to
compute it with the replica trick, in terms of derivatives of the on-shell
Euclidean action. Then, we explain the generalization of the Euler theorem to
4D conically singular manifolds without $U\left(  1\right)  $ isometry.
Finally, we consider the renormalized Euclidean gravitational E-H action, as
obtained by the Kounterterms procedure, and comment on its properties and
usefulness for the computation of $S_{EE}^{ren}$.

\subsection{Entanglement Entropy and replica trick\label{EE_Replica}}

The EE \cite{RT1}-\cite{Maldacena1} is defined as the von Neumann entropy of
the reduced density matrix of a quantum subsystem $A$, i.e.%

\begin{equation}
S_{EE}=-Tr\left(  \widehat{\rho}_{A}\ln\widehat{\rho}_{A}\right)  ,
\label{VonNewmann}%
\end{equation}
and it encodes the degree of entanglement of the subsystem $A$ with the rest
of the system ($A^{c}$). The first proposal for computing EEs of CFTs in the
AdS/CFT framework was the RT formula \cite{RT1}. Said formula states that the
EE of an entangling region $A$ in a (D-1)-dimensional CFT with a D-dimensional
AAdS gravity dual is equal to the volume of a codimension 2 minimal
hypersurface ($\Sigma$) in the AAdS bulk whose border is conformal to the one
of the entangling region ($A$) at the conformal boundary; i.e., $S_{EE}%
=\frac{Vol(\Sigma)}{4G}$ (in natural units). This formula is analogous to the
Bekenstein-Hawking entropy formula for a black hole \cite{BHE1}-\cite{BHE3},
and it was shown (e.g., by Lewkowycz and Maldacena in \cite{Maldacena1}) that
indeed both formulas can be obtained from the replica trick \cite{EECovariant}%
-\cite{Marika}.

The computation of the EE by the replica trick considers that
eq.(\ref{VonNewmann}) can be re-written as%

\begin{equation}
S_{EE}=\lim_{n\rightarrow1}-\frac{1}{n-1}\ln(Tr\left(  \widehat{\rho}_{A}%
^{n}\right)  ), \label{RenyiConst}%
\end{equation}
and therefore, the EE is expressed in terms of the trace of the n-th power of
the reduced density matrix. In order to compute this trace, one constructs a
branched cover of the conformal boundary $C$ where the CFT is defined. This is
achived by gluing together n copies of the original (Euclideanized) boundary
with a cut along the entangling region whose EE is being computed
\cite{RenyiXiDong}. The gluing is done such that, when defining an angular
coordinate that circles around the border of the entangling region, after
$2\pi$ rotations along the coordinate, the copies are cyclically permuted.
Labelling this branched cover manifold as $C_{n}$, one realizes that it has a
$Z_{n}$ symmetry corresponding to cycling from one copy (replica) of the CFT
to another. Finally, one defines the orbifold $\widehat{C}_{n}$ as the
quotient of the cover manifold by the permutation symmetry, i.e., $\widehat
{C}_{n}=C_{n}/Z_{n}$. The orbifold $\widehat{C}_{n}$ is conically singular,
with an opening angle of $\frac{2\pi}{n}$. However, because the permutation
symmetry is the discrete symmetry $Z_{n}$, the orbifold does not have a
$U\left(  1\right)  $ isometry in general.

Considering the $\widehat{C}_{n}$ orbifold, $Tr\left(  \widehat{\rho}_{A}%
^{n}\right)  $ can be computed in terms of the partition function of the
replica CFT defined on the orbifold, as%

\begin{equation}
Tr\left(  \widehat{\rho}_{A}^{n}\right)  =n\left(  \ln\left(  Z\left(
\widehat{C}_{n}\right)  \right)  -\ln\left(  Z\left(  \widehat{C}_{1}\right)
\right)  \right)  , \label{TraceReplica}%
\end{equation}
where $\widehat{C}_{1}$ (which is equal to $C$) is the manifold of the
original CFT and $Z\left(  \widehat{C}_{1}\right)  $ is its partition function.

One then defines the orbifold $\widehat{M}_{n}$ as the extension of
$\widehat{C}_{n}$ into the AAdS bulk, by requiring the bulk metric to be a
solution of the equations of motion. Because the orbifold $\widehat{M}_{n}$ is
a solution in the bulk, the semi-classical approximation can be used to write
the partition functions in eq(\ref{TraceReplica}) in terms of the
corresponding gravitational Euclidean on-shell actions in the AAdS bulk
(including boundary terms). Then, in the saddle-point approximation, one has
that $\ln(Z(\widehat{C}_{n}))=-I_{E}(\widehat{M}_{n})$, and therefore,%

\begin{equation}
Tr\left(  \widehat{\rho}_{A}^{n}\right)  =-n\left(  I_{E}(\widehat{M}%
_{n})-I_{E}(\widehat{M}_{1})\right)  .
\end{equation}
Thus, in the AdS/CFT context, the EE computed by the replica trick can be
written as%

\begin{equation}
S_{EE}=\lim_{n\rightarrow1}\frac{n}{n-1}(I_{E}(\widehat{M}_{n})-I_{E}%
(\widehat{M}_{1}))=\left.  n^{2}\partial_{n}I_{E}\left(  \widehat{M}%
_{n}\right)  \right\vert _{n=1}. \label{Replica_EE}%
\end{equation}
Finally, for ease of computation, we define the angular parameter $\alpha$
such that $\alpha=\frac{1}{n},$ where the cone then has an angular deficit
given by $2\pi\left(  1-\alpha\right)  =2\pi\left(  1-\frac{1}{n}\right)  $.
Thus, in the case of a 3D CFT with a 4D AAdS gravity dual, the EE is%

\begin{equation}
S_{EE}=\left.  -\partial_{\alpha}I_{E}\left(  \widehat{M}_{4}^{(\alpha
)}\right)  \right\vert _{\alpha=1}, \label{Replica_Alpha_EE}%
\end{equation}
where now $\widehat{M}_{4}^{(\alpha)}$ denotes the 4D orbifold with angular
deficit given by $2\pi\left(  1-\alpha\right)  $.

In order to evaluate this Euclidean action, we first need to discuss some
properties of differential geometry in conically singular manifolds
\cite{Solodukhin1}-\cite{Cone3}. In particular, in the next section, we review
a generalization of the Euler theorem for squashed cones (conically singular
manifolds without $U\left(  1\right)  $ isometry) in 4D.

\subsection{Euler theorem for conically singular manifolds in 4D}

In differential geometry, topological invariants are interesting because they
characterize properties of manifolds that are robust under continuous
deformations of their metric. For example, the Euler characteristic in $D=2m$
can be written as the integral of a precise combination of a product of
$m-$curvature terms, with the addition of the $m-$th Chern form in a manifold
with boundaries. The Chern form is expressible considering both intrinsic and
extrinsic curvatures of the boundary's induced metric. Therefore, this way of
writing the Euler characteristic provides a global relation between the
curvature of a bulk manifold, and the curvatures of its boundary. In
particular, the Euler theorem \cite{K1Even}, which is valid for $2m-$%
dimensional manifolds, states that%

\begin{equation}
\int\limits_{M_{2m}}\varepsilon_{2m}=\left(  4\pi\right)  ^{m}m!\chi\left(
M_{2m}\right)  +\int\limits_{\partial M_{2m}}B_{2m-1}, \label{Euler_Theo}%
\end{equation}
where $\varepsilon_{2m}$ is the Euler density in $2m$ dimensions, $\chi\left(
M_{2m}\right)  $ is the Euler characteristic of the manifold $M_{2m}$, and
$B_{2m-1}$ is the $m-$th Chern form at the boundary of the manifold. In the
particular case of $m=2$, and therefore $\dim\left(  M_{2m}\right)  =4$, the
Euler density $\varepsilon_{4}$ is the usual Gauss-Bonnet term and $B_{3}$ is
the second Chern form (given in eq.(\ref{B_3}) in Gauss normal coordinates).

As we will see in section \ref{Renorm}, in order to obtain the renormalized
version of $S_{EE}$, we need to evaluate either $\varepsilon_{4}$ or $B_{3}$
on conically singular manifolds (without $U\left(  1\right)  $ rotational
isometry). To this end, we consider the results obtained by Fursaev, Patrushev
and Solodukhin (FPS) \cite{SolodukhinNew} regarding the computation of
quadratic terms in the curvature, for conically singular manifolds in 4D.

In order to compute the integral of the quadratic terms, which correspond to
the Ricci scalar squared, the Ricci tensor squared and the Riemann tensor
squared, FPS used the methods of distributional geometry, as described in
\cite{Solodukhin1}\cite{SolodukhinNew}. There, a conically singular orbifold
was considered as the limit of a sequence of regular manifolds whose metrics
are parametrized by a certain regularization parameter. Then, the quadratic
terms are computed, and the parameter is taken to zero, in order to recover
the conically singular manifold (for further details, we refer the reader to
the original papers).

In particular, FPS obtained that the integral of the square of the Riemann
tensor evaluated on the 4D orbifold $\widehat{M}_{4}^{\left(  \alpha\right)
}$ is given by%

\begin{equation}%
\begin{array}
[c]{c}%
{\displaystyle\int\limits_{\widehat{M}_{4}^{\left(  \alpha\right)  }}}
d^{4}x\sqrt{G}\left(  R^{\left(  \alpha\right)  }\right)  _{\nu\sigma\lambda
}^{\mu}\left(  R^{\left(  \alpha\right)  }\right)  _{\mu}^{\nu\sigma\lambda}=%
{\displaystyle\int\limits_{M_{4}}}
d^{4}x\sqrt{G}\left(  R^{\left(  r\right)  }\right)  _{\nu\sigma\lambda}^{\mu
}\left(  R^{\left(  r\right)  }\right)  _{\mu}^{\nu\sigma\lambda}+\\
8\pi\left(  1-\alpha\right)
{\displaystyle\int\limits_{\Sigma}}
d^{2}x\sqrt{\gamma}\left(  R_{\left(  i\right)  \left(  j\right)  \left(
i\right)  \left(  j\right)  }^{\left(  r\right)  }-\left(  K_{\left(
i\right)  }\right)  _{b}^{a}\left(  K_{\left(  i\right)  }\right)  _{a}%
^{b}\right)  +O\left(  \left(  1-\alpha\right)  ^{2}\right)  ,
\end{array}
\label{Riemann}%
\end{equation}
where $\left(  R^{\left(  \alpha\right)  }\right)  _{\nu\sigma\lambda}^{\mu}$
denotes the bulk Riemann tensor evaluated on the orbifold, $\left(  R^{\left(
r\right)  }\right)  _{\nu\sigma\lambda}^{\mu}$ represents the regular part of
the bulk Riemann tensor, $M_{4}$ refers to the regular manifold given in the
$\alpha\rightarrow1$ limit (where $2\pi\left(  1-\alpha\right)  $ is the
angular deficit of the cone), $G_{\mu\nu}$ corresponds to the bulk metric of
the manifold, $\Sigma$ is the codimension-2 surface located at the tip of the
cone and given by the fixed-point set of the $Z_{n}$ symmetry of the orbifold,
$\gamma_{ab}$ is the induced metric on $\Sigma$, $R_{\left(  i\right)  \left(
j\right)  \left(  i\right)  \left(  j\right)  }^{\left(  r\right)  }$ denotes
the corresponding components of the Riemann tensor where $\left(  i\right)  $
and $\left(  j\right)  $ are the indices of the foliation ($i,j=1,2$) and
$\left(  K_{\left(  i\right)  }\right)  _{b}^{a}$ is the extrinsic curvature
tensor of the surface $\Sigma$ with respect to the $i-$th direction of the
foliation that is normal to the surface ($i=1,2$), where a sum over repeated
foliation indices is implied. We note that$\left(  R^{\left(  \alpha\right)
}\right)  _{\alpha\beta}^{\mu\nu}$ is divergent at $\Sigma$, and it can be
written as%

\begin{equation}%
\begin{tabular}
[c]{l}%
$\left(  R^{\left(  \alpha\right)  }\right)  _{\sigma\lambda}^{\mu\nu}=\left(
R^{\left(  r\right)  }\right)  _{\sigma\lambda}^{\mu\nu}+2\pi\left(
1-\alpha\right)  \left(  N_{\sigma\lambda}^{\mu\nu}+T_{\sigma\lambda}^{\mu\nu
}\right)  \delta_{\Sigma},$\\
$N_{\rho\lambda}^{\mu\nu}=\left[  \left(  n_{\left(  i\right)  }\right)
^{\mu}\left(  n_{\left(  i\right)  }\right)  _{\sigma}\left(  n_{\left(
j\right)  }\right)  ^{\nu}\left(  n_{\left(  j\right)  }\right)  _{\lambda
}-\left(  n_{\left(  i\right)  }\right)  ^{\mu}\left(  n_{\left(  i\right)
}\right)  _{\lambda}\left(  n_{\left(  j\right)  }\right)  ^{\nu}\left(
n_{\left(  j\right)  }\right)  _{\sigma}\right]  ,$%
\end{tabular}
\ \ \label{ConicRiemann}%
\end{equation}
where $\delta_{\Sigma}$ is a codimension-2 delta function which only has
support on $\Sigma$, $\left(  n_{\left(  i\right)  }\right)  _{\mu}$ is the
$i-$th normal vector to the $\Sigma$ surface ($i=1,2$) and $T_{\sigma\lambda
}^{\mu\nu}$ is a tensor which depends on the extrinsic curvatures of $\Sigma$
with respect to the two directions of the foliation. In the case that the cone
has a $U\left(  1\right)  $ rotational symmetry, $T_{\sigma\lambda}^{\mu\nu
}=0$. However in our case, although $T_{\sigma\lambda}^{\mu\nu}$ is left
unspecified, it does encode the extrinsic curvature contributions to the
quadratic terms.

Analogously, FPS obtained that for the square of the Ricci tensor,%

\begin{equation}%
\begin{array}
[c]{c}%
{\displaystyle\int\limits_{\widehat{M}_{4}^{\left(  \alpha\right)  }}}
d^{4}x\sqrt{G}\left(  R^{\left(  \alpha\right)  }\right)  _{\mu\nu}\left(
R^{\left(  \alpha\right)  }\right)  ^{\mu\nu}=%
{\displaystyle\int\limits_{M_{4}}}
d^{4}x\sqrt{G}\left(  R^{\left(  r\right)  }\right)  _{\mu\nu}\left(
R^{\left(  r\right)  }\right)  ^{\mu\nu}+\\
4\pi\left(  1-\alpha\right)
{\displaystyle\int\limits_{\Sigma}}
d^{2}x\sqrt{\gamma}\left(  R_{\left(  i\right)  \left(  i\right)  }^{\left(
r\right)  }-\frac{1}{2}\left(  K_{\left(  i\right)  }\right)  _{a}^{a}\left(
K_{\left(  i\right)  }\right)  _{b}^{b}\right)  +O\left(  \left(
1-\alpha\right)  ^{2}\right)  ,
\end{array}
\label{Ricci}%
\end{equation}
and for the square of the Ricci scalar,%

\begin{equation}%
{\displaystyle\int\limits_{\widehat{M}_{4}^{\left(  \alpha\right)  }}}
d^{4}x\sqrt{G}\left(  R^{\left(  \alpha\right)  }\right)  ^{2}=%
{\displaystyle\int\limits_{M_{4}}}
d^{4}x\sqrt{G}\left(  R^{\left(  r\right)  }\right)  ^{2}+8\pi\left(
1-\alpha\right)
{\displaystyle\int\limits_{\Sigma}}
d^{2}x\sqrt{\gamma}\left(  R^{\left(  r\right)  }\right)  +O\left(  \left(
1-\alpha\right)  ^{2}\right)  . \label{Ricci_Scalar}%
\end{equation}
Because in the computation of $S_{EE}$ we need to take the $\alpha
\rightarrow1$ limit, it is safe to neglect terms of quadratic or higher order
in $\left(  1-\alpha\right)  $.

Finally, we have that the Gauss-Codazzi decomposition of the regular part of
the Ricci scalar on $M_{4}$ gives
\begin{equation}
R^{\left(  r\right)  }=-R_{\left(  i\right)  \left(  j\right)  \left(
i\right)  \left(  j\right)  }^{\left(  r\right)  }+2R_{\left(  i\right)
\left(  i\right)  }^{\left(  r\right)  }+\mathcal{R}-\left(  K_{\left(
i\right)  }\right)  _{a}^{a}\left(  K_{\left(  i\right)  }\right)  _{b}%
^{b}+\left(  K_{\left(  i\right)  }\right)  _{a}^{b}\left(  K_{\left(
i\right)  }\right)  _{b}^{a}, \label{Gauss-Codazzi_R}%
\end{equation}
where we $\mathcal{R}$ is the intrinsic Ricci scalar at the surface $\Sigma$
(computed with the induced metric $\gamma_{ab}$), and the other quantities
have the same meaning as for the quadratic terms presented above.

Now, considering eqs.(\ref{Riemann}-\ref{Gauss-Codazzi_R}), we evaluate
$\varepsilon_{4}$ on the $\widehat{M}_{4}^{\left(  \alpha\right)  }$ orbifold.
The Gauss-Bonnet term is given by%

\begin{equation}
\varepsilon_{4}=\sqrt{G}d^{4}x\left(  R_{\nu\sigma\lambda}^{\mu}R_{\mu}%
^{\nu\sigma\lambda}-4R_{\mu\nu}R^{\mu\nu}+R^{2}\right)  , \label{Gauss-Bonnet}%
\end{equation}
and therefore, we obtain that%

\begin{equation}
\int\limits_{\widehat{M}_{4}^{\left(  \alpha\right)  }}\varepsilon
_{4}^{\left(  \alpha\right)  }=\int\limits_{M_{4}}\varepsilon_{4}^{\left(
r\right)  }+8\pi\left(  1-\alpha\right)  \int\limits_{\Sigma}\varepsilon
_{2}+O\left(  \left(  1-\alpha\right)  ^{2}\right)  , \label{Euler_conical.}%
\end{equation}
where we used that $\varepsilon_{2}=\mathcal{R}\sqrt{\gamma}d^{2}x$ is the
usual 2D Gauss-Bonnet term, which depends on the intrinsic Ricci scalar at the
surface $\Sigma$.

Furthermore, considering that (as shown in FPS) for squashed-cone manifolds in
4D, the Euler characteristic obeys the relation%

\begin{equation}
\chi_{4}\left(  \widehat{M}_{4}^{\left(  \alpha\right)  }\right)  =\chi
_{4}\left(  M_{4}\right)  +\left(  1-\alpha\right)  \chi_{2}\left(
\Sigma\right)  +O\left(  \left(  1-\alpha\right)  ^{2}\right)  ,
\label{Euler_Char}%
\end{equation}
and also using eq.(\ref{Euler_Theo}) for the $m=1$ and $m=2$ cases, we obtain
that the boundary terms (given by the corresponding Chern forms) are related by%

\begin{equation}
\int\limits_{\partial\widehat{M}_{4}^{\left(  \alpha\right)  }}B_{3}^{\left(
\alpha\right)  }=\int\limits_{\partial M_{4}}B_{3}^{\left(  r\right)  }%
+8\pi\left(  1-\alpha\right)  \int\limits_{\partial\Sigma}B_{1}+O\left(
\left(  1-\alpha\right)  ^{2}\right)  , \label{Euler_Boundary_Conical}%
\end{equation}
where $B_{1}$ is evaluated at the boundary of the codimension-2 surface
$\Sigma$.

It is precisely this last relation which will be used in section \ref{Renorm},
in order to evaluate the Euclidean action in the orbifold, which will
ultimately give the expression for the renormalized EE when considering the
renormalized Euclidean action which will be discussed in the following subsection.

\subsection{Renormalized Euclidean action and Topological Invariants}

In order to obtain a renormalized version of eq.(\ref{Replica_Alpha_EE}), to
be able to compute the finite part of the EE, we need to consider a suitably
renormalized Euclidean action for the bulk gravity theory.

For AAdS spacetimes, there are different prescriptions for renormalizing the
Euclidean on-shell action. The standard Holographic Renormalization method
consists on adding counterterms to the action as surface terms,
\cite{UsualCounterJohnson}-\cite{SkenderisAndPapa2}. In doing so, the
divergences occuring due to the presence of the infinite conformal factor in
the metric at the boundary, as seen in its Fefferman-Graham expansion
\cite{FeffermanGraham} are cancelled out. The counterterms are functionals of
the boundary metric, its intrinsic curvature and covariant derivatives
thereof, in order to be consistent with a well posed variational principle for
the conformal class of spacetimes $\left[  h_{ij}\right]  $ at the boundary
\cite{SkenderisAndPapa1}\cite{SkenderisAndPapa2}, after the
Gibbons-Hawking-York term is included. Although there is a systematic
procedure for computing the counterterms, in principle, at any order in the
holographic radial coordinate $\rho$ and for any number of spacetime
dimensions \cite{DirichletKraus(quasiloc)}, the number of counterterms
required grows rapidly with the dimension. Furthermore, the functional form of
the terms in the series is different for different gravity actions including
higher-curvature theories (e.g., Lovelock gravity theories).

The Kounterterms procedure, developed in ref.\cite{K1Even}, and further
understood in ref.\cite{KounterComparison2}, consists on adding a given
boundary term to the AdS gravity action in order to both attain a well defined
variational principle and to render the action finite. The particular term
that is added is universal for all gravity theories of Lovelock type and
depends only on the number of dimensions of spacetime and on whether said
number is odd or even. In the case of AAdS spacetimes in even dimensions, with
$D=2m$, the term added is the $m-$th Chern form \cite{K1Even}. For
odd-dimensional spacetimes, the term added corresponds to the boundary term of
the Chern-Simons transgression form of the AdS group \cite{K2Odd}. In both
cases, the added boundary term depends on both the intrinsic and extrinsic
($K_{j}^{i}$) curvatures of the boundary in the radial foliation of the
spacetime, and hence the name Kounterterms. Therefore, it is easy to
particularize to the case of Fefferman-Graham expansion (with respect to the
holographic radial coordinate $\rho$). In even-dimensional manifolds, there is
a relation between the added boundary terms and topological terms. Indeed, the
$m-$th Chern form is the boundary term associated with the Euler theorem,
which relates the integral of the Euler term in the bulk with the Euler
characteristic of the manifold, as mentioned in the previous subsection.

It is important to note that in \cite{KounterComparison}%
\cite{KounterComparison2}, it was proven that the addition of Kounterterms is
equivalent to the standard Holographc Renormalization procedure in Einstein
gravity in even dimmensions. As a matter of fact, standard counterterms are
recovered when expressing these extrinsic counterterms in terms of the
intrinsic curvature at the boundary (making extensive use of the FG expansion
for $K_{j}^{i}$ ), order by order in the holographic radial coordinate $\rho$.
However, the universality of the Kounterterms method with respect to different
gravity theories (i.e., all theories of Lovelock type) and the fact that the
closed-form expression for the boundary term is known for any dimension are
its main practical advantages over the standard Holographic Renormalization
procedure; but also its relation to topology is interesting on its own.

In this paper, we consider the renormalized action given by the Kounterterms
prescription, which for the case of EH gravity in 4D AAdS manifolds is given
by \cite{K1Even}%

\begin{equation}
I_{E}^{ren}=\frac{1}{16\pi G}\left(
{\displaystyle\int\limits_{M_{4}}}
d^{4}x\sqrt{G}\left(  R-2\Lambda\right)  +\frac{\ell^{2}}{4}%
{\displaystyle\int\limits_{\partial M_{4}}}
B_{3}\right)  , \label{I^ren_4D}%
\end{equation}
where $\Lambda=-\frac{3}{\ell^{2}}$, and $B_{3}$ is given by%

\begin{equation}
B_{3}=-4\int_{0}^{1}dtd^{3}x\sqrt{h}\delta_{\lbrack i_{1}i_{2}i_{3}]}%
^{[j_{1}j_{2}j_{3}]}K_{j_{1}}^{i_{1}}\left(  \frac{1}{2}\mathcal{R}%
_{j_{2}j_{3}}^{i_{2}i_{3}}-t^{2}K_{j_{2}}^{i_{2}}K_{j_{3}}^{i_{3}}\right)  .
\label{B_3}%
\end{equation}

In eq.(\ref{B_3}), $h_{ij}$ is the metric at the boundary of spacetime,
$\mathcal{R}_{ij}^{k\ell}$ is the Riemann curvature tensor at the boundary,
computed with the $h_{ij}$ metric, and $K_{j}^{i}$ is the extrinsic curvature
tensor of the boundary with respect to a radial foliation along the
holographic radial coordinate $\rho$. The main reason for adopting this
renormalization scheme is that the boundary term $B_{3}$ can be directly
evaluated in the orbifold $\widehat{M}_{4}^{\left(  \alpha\right)  }$ using
the generalized Euler theorem for the boundary terms, as presented in
eq.(\ref{Euler_Boundary_Conical}). Then, the counterterm for the renormalized
EE directly becomes the $B_{1}$ term evaluated at the entangling surface
$\partial\Sigma$ as shown in eq.(\ref{S_EE^ren}). This will be explained in
detail in the following section.

\section{Renormalization of EE in AdS$_{4}$/CFT$_{3}$ through the Chern
form\label{Renorm}}

After introducing the renormalized Euclidean action for the dual gravitational
theory and also the generalization of the Euler theorem for 4D squashed-cone
manifolds, we proceed to compute the renormalized EE $\left(  S_{EE}%
^{ren}\right)  $ by means of the replica trick. This is done by evaluating
eq.(\ref{Replica_Alpha_EE}) using the renormalized gravitational action of
eq.(\ref{I^ren_4D}). This assumes that if the gravitational action is itself
renormalized, then the resulting EE will be renormalized as well.

The renormalized Euclidean on-shell action, evaluated on the conically
singular manifold $\widehat{M}_{4}^{(\alpha)}$, is given by%

\begin{equation}
I_{E}^{ren}=\frac{1}{16\pi G}\left(
{\displaystyle\int\limits_{\hat{M}_{4}^{\left(  \alpha\right)  }}}
d^{4}x\sqrt{G}\left(  R^{\left(  \alpha\right)  }-2\Lambda\right)  +\frac
{\ell^{2}}{4}%
{\displaystyle\int\limits_{\partial\hat{M}_{4}^{\left(  \alpha\right)  }}}
B_{3}^{\left(  \alpha\right)  }\right)  . \label{I_ren_alpha}%
\end{equation}
As it was discussed by Lewkowicz and Maldacena \cite{Maldacena1} and by Dong
\cite{RenyiXiDong}, the Einstein-Hilbert part of eq.(\ref{I_ren_alpha}) gives
the usual RT area formula for the EE, when computing the derivative of
eq.(\ref{Replica_Alpha_EE}) with respect to the conical angle parameter
$\alpha$. Therefore, the counterterm that regularizes the EE will come from
the $B_{3}^{\left(  \alpha\right)  }$ part. We defne the counterterm of the
Euclidean action as%

\begin{equation}
I_{E}^{ct}=\frac{\ell^{2}}{64\pi G}%
{\displaystyle\int\limits_{\partial\hat{M}_{4}^{\left(  \alpha\right)  }}}
B_{3}^{\left(  \alpha\right)  }, \label{I_ct}%
\end{equation}
and therefore, we proceed to compute the counterterm of the EE ($S_{EE}^{ct}$) as%

\begin{equation}
S_{EE}^{ct}=\left.  -\partial_{\alpha}I_{E}^{ct}\left(  \partial\widehat
{M}_{4}^{(\alpha)}\right)  \right\vert _{\alpha=1}, \label{S_ct}%
\end{equation}
such that $S_{EE}^{ren}=S_{EE}^{RT}+S_{EE}^{ct}$, where $S_{EE}^{RT}$ is the
usual RT prescription for the EE.

Using eq.(\ref{Euler_Boundary_Conical}) to evaluate $I_{E}^{ct}$, we have that%

\begin{equation}
S_{EE}^{ct}=\frac{\ell^{2}}{8G}\int\limits_{\partial\Sigma}B_{1},
\label{S_EE^ct}%
\end{equation}
and therefore, we recover the expression for $S_{EE}^{ren}$ given in
eq.(\ref{S_EE^ren}).

In the next subsections, we will expand the integrands of $S_{EE}^{ren}$ in
their corresponding FG expansions, in order to verify the finiteness of the
renormalized EE, and also in order to show that our result is equivalent to
the one obtained in \cite{Marika}. We will also compute $S_{EE}^{ren}$ for the
particular case of a disk-like entangling region in the 3D CFT, with a global
AdS$_{4}$ gravitational dual (corresponding to the ground state of the CFT).
In order to do this, we will consider the explicit embedding of the minimal
surface $\Sigma$ and its boundary $\partial\Sigma$, as given in
\cite{HungMyersSmolkin}\cite{SchwimmerAndTheisen} and as explained in what follows.

\subsection{Explicit covariant embedding}

Following the works by Hung, Myers and Smolkin \cite{HungMyersSmolkin}, and by
Schwimmer and Theisen \cite{SchwimmerAndTheisen} we consider that the
embedding of the minimal surface $\Sigma$ on the bulk is given by%

\begin{equation}%
\begin{tabular}
[c]{l}%
$x^{i}\left(  \tau,y^{a}\right)  =\left(  x^{\left(  0\right)  }\right)
^{i}\left(  y^{a}\right)  +\tau\left(  x^{\left(  2\right)  }\right)
^{i}\left(  y^{a}\right)  +...$\\
$\left(  x^{\left(  2\right)  }\right)  ^{i}\left(  y^{a}\right)  =\frac
{\ell^{2}}{2\left(  d-2\right)  }\kappa^{i}\left(  y^{a}\right)  ,$%
\end{tabular}
\ \ \label{Sigma_embedding}%
\end{equation}
where $\left\{  \rho,x^{i}\right\}  $ are bulk coordinates and $\left\{
\tau,y^{a}\right\}  $ are coordinates on the worldvolume of $\Sigma$.

We label the AAdS bulk metric as $G_{\mu\nu}$ and the metric at the spacetime
boundary as $h_{ij}$. Analogously, we label the induced metric on $\Sigma$ as
$\gamma_{ab}$ and the induced metric on its boundary ($\partial\Sigma$) as
$\widetilde{\gamma}_{ab}$. As it is well known (see, e.g., \cite{Imbimbo}),
the metric $G_{\mu\nu}$ has a FG expansion \cite{FeffermanGraham} given by%

\begin{equation}%
\begin{tabular}
[c]{l}%
$ds_{G}^{2}=G_{\mu\nu}dx^{\mu}dx^{\nu}=\frac{\ell^{2}d\rho^{2}}{4\rho^{2}%
}+h_{ij}\left(  \rho,x\right)  dx^{i}dx^{j},$\\
$h_{ij}\left(  \rho,x\right)  =\frac{g_{ij}\left(  \rho,x\right)  }{\rho},$\\
$g_{ij}\left(  \rho,x\right)  =g_{ij}^{\left(  0\right)  }\left(  x\right)
+\rho g_{ij}^{\left(  2\right)  }\left(  x\right)  +...,$%
\end{tabular}
\ \label{F-G_BulkMetric}%
\end{equation}
where $\rho$ is the holographic radial coordinate (the spacetime boundary is
located at $\rho=0$).

Now, the induced metric $\gamma_{ab}$ is defined as%

\begin{equation}
\gamma_{ab}=\frac{\partial x^{\mu}}{\partial y^{a}}\frac{\partial x^{\nu}%
}{\partial y^{b}}G_{\mu\nu}, \label{InducedMetricSigma}%
\end{equation}
and upon choosing the diffeomorphism gauge as $\tau=\rho$ and $\gamma_{a\tau
}=0$, one obtains that%

\begin{equation}%
\begin{tabular}
[c]{l}%
$ds_{\gamma}^{2}=\gamma_{ab}dy^{a}dy^{b}=\frac{\ell^{2}}{4\tau^{2}}\left(
1+\frac{\tau\ell^{2}}{\left(  d-2\right)  ^{2}}\kappa^{i}\kappa^{j}%
g_{ij}^{\left(  0\right)  }+...\right)  d\tau^{2}+\widetilde{\gamma}%
_{ab}\left(  \tau,y\right)  dy^{a}dy^{b},$\\
$\widetilde{\gamma}_{ab}\left(  \tau,y\right)  =\frac{\sigma_{ab}\left(
\tau,y\right)  }{\tau},$\\
$\sigma_{ab}\left(  \tau,y\right)  =\sigma_{ab}^{\left(  0\right)  }\left(
y\right)  +\tau\sigma_{ab}^{\left(  2\right)  }\left(  y\right)  +...,$%
\end{tabular}
\ \ \label{F-G_SigmaMetric}%
\end{equation}
which has the form of a FG-like expansion for the induced metric on $\Sigma$.
To make this last statement more precise, $\gamma_{ab}$ is given by the
pullback onto $\Sigma$ of the $G_{\mu\nu}$ bulk metric in the FG gauge.

We now explain the meaning of the coefficients in the FG expansions of the
previous equations. In particular, $d$ is the dimension of the boundary of
spacetime ($d=3$ in our case), $g_{ij}^{\left(  0\right)  }$ is the metric at
the conformal boundary (where the 3D CFT is defined) and $\sigma_{ab}^{\left(
0\right)  }$ is the induced metric on the entangling surface $\partial A$
(embedded in the conformal boundary) which is given by
\begin{equation}
\sigma_{ab}^{\left(  0\right)  }=\frac{\partial\left(  x^{\left(  0\right)
}\right)  ^{i}}{\partial y^{a}}\frac{\partial\left(  x^{\left(  0\right)
}\right)  ^{j}}{\partial y^{b}}g_{ij}^{\left(  0\right)  }.
\end{equation}
Furthermore, $g_{ij}^{\left(  2\right)  }=-\ell^{2}S_{ij}^{\left(  0\right)
}$ where $S_{ij}^{\left(  0\right)  }$ is the Schouten tensor of the
$g_{ij}^{\left(  0\right)  }$ metric given by
\begin{equation}
\left(  S^{\left(  0\right)  }\right)  _{ij}=\frac{1}{\left(  d-2\right)
}\left(  R_{ij}^{\left(  0\right)  }-\frac{g_{ij}^{\left(  0\right)  }%
}{2\left(  d-1\right)  }R^{\left(  0\right)  }\right)  ,
\end{equation}
and $\sigma_{ab}^{\left(  2\right)  }$ is given by
\begin{equation}
\sigma_{ab}^{\left(  2\right)  }=\frac{\partial\left(  x^{\left(  0\right)
}\right)  ^{i}}{\partial y^{a}}\frac{\partial\left(  x^{\left(  0\right)
}\right)  ^{j}}{\partial y^{b}}g_{ij}^{\left(  2\right)  }-\frac{\ell^{2}%
}{\left(  d-2\right)  }\kappa^{i}\kappa_{ab}^{j}\left(  g^{\left(  0\right)
}\right)  _{ij}.
\end{equation}
Finally,
\begin{equation}
\kappa^{i}=\hat{n}_{\left(  n\right)  }^{i}\kappa_{ab}^{\left(  n\right)
}\left(  \sigma^{\left(  0\right)  }\right)  ^{ab},
\end{equation}
where the extrinsic curvatures $\kappa_{ab}^{\left(  n\right)  }$ are defined
with respect to the foliation that is normal to $\partial A$ (which is
conformal to $\partial\Sigma$) embedded in the conformal boundary (which is
conformal to the boundary of spacetime), and $\hat{n}_{\left(  n\right)  }%
^{i}$ are the vectors along the directions of the foliation ($n=1,2$).

Having introduced the explicit covariant embedding of $\Sigma$ in $\widehat
{M}_{4}^{(\alpha)}$, and the corresponding FG expansions of $G_{\mu\nu}$ and
$\gamma_{ab}$, we now proceed to show that $S_{EE}^{ren}$, as defined in
eq.(\ref{S_EE^ren}), is finite and equivalent to the standard expression given
in eq.(\ref{S_ren_usual_4D}), and first obtained in \cite{Marika}.

\subsection{Proof of finiteness of $S_{EE}^{ren}$\label{FinitenessProof}}

With the previously considered embedding, we can check the cancellation of
divergences in $S_{EE}^{ren}$ for 3D CFTs with 4D AAdS gravity duals. We note
that the explicit value of $S_{EE}^{ren}$ depends on the shape of the
entangling surface $\partial A$ at the conformal boundary. Here, we simply
exhibit the divergences in $S_{EE}^{RT}$ (the standard Ryu-Takayanagi EE) and
check that they are exactly cancelled by $S_{EE}^{ct}$ (the EE counterterm
given in eq.(\ref{S_EE^ct})). This cancellation is independent of the shape of
$\partial A$.

In particular, we have%

\begin{equation}%
\begin{tabular}
[c]{l}%
$B_{1}=-2dy\sqrt{\widetilde{\gamma}}\delta_{a}^{b}k_{b}^{a},$\\
$S_{EE}^{ct}=-\frac{\ell^{2}}{4G}\int\limits_{\partial\Sigma}dy\sqrt
{\widetilde{\gamma}}\widetilde{\gamma}^{ab}k_{ab},$\\
$S_{EE}^{RT}=\frac{1}{4G}\int\limits_{\Sigma}d^{2}y\sqrt{\gamma},$%
\end{tabular}
\ \ \ \label{4D_EE_terms}%
\end{equation}
where $k_{ab}$ is the extrinsic curvature of $\partial\Sigma$ with respect to
the radial foliation along the holographic radial coordinate $\rho$ (not to be
confused with $\kappa_{ab}^{\left(  n\right)  }$ for $\partial A$, or with
$K_{ij}$\ for $B$).

Now, we consider the FG expansion of each of the pieces. From
eq.(\ref{F-G_SigmaMetric}), we have that the square root of the determinant of
the metric on $\Sigma$ and on $\partial\Sigma$ are given, respectively, by%

\begin{equation}%
\begin{tabular}
[c]{l}%
$\sqrt{\gamma}=\frac{\ell\sqrt{\sigma^{\left(  0\right)  }}}{2\rho^{d/2}%
}\left(  1+\rho\left[  \frac{\ell^{2}}{2\left(  d-2\right)  ^{2}}\kappa
^{i}\kappa^{j}g_{ij}^{\left(  0\right)  }+\frac{1}{2}tr[\sigma^{\left(
2\right)  }]\right]  +...\right)  ,$\\
$\sqrt{\widetilde{\gamma}}=\frac{\sqrt{\sigma^{\left(  0\right)  }}}%
{\rho^{\left(  d-2\right)  /2}}\left(  1+\frac{\rho}{2}tr[\sigma^{\left(
2\right)  }]+...\right)  ,$%
\end{tabular}
\ \ \ \label{MetricDets}%
\end{equation}
where $tr[\sigma^{\left(  2\right)  }]$ denotes the trace of the $\sigma
_{ab}^{\left(  2\right)  }$ tensor, given in the paragraph following
eq.(\ref{F-G_SigmaMetric}). Also, $\widetilde{\gamma}^{ab}$ is the inverse of
the induced metric on $\partial\Sigma$, and it is given by $\widetilde{\gamma
}^{ab}=\rho\left(  \left(  \sigma^{\left(  0\right)  }\right)  ^{ab}%
-\rho\left(  \sigma^{\left(  2\right)  }\right)  ^{ab}+...\right)  $. Now,
considering the FG-like expansion of the induced metric on $\Sigma$, the
extrinsic curvature $k_{ab}$ with respect to the radial foliation is computed,
by definition, as $k_{ab}=\frac{-1}{2\sqrt{\gamma_{\rho\rho}}}\partial_{\rho
}\widetilde{\gamma}_{ab}$. Thus, we have that%

\begin{equation}
k_{ab}=\frac{\sigma_{ab}^{\left(  0\right)  }}{\ell\rho}\left(  1-\frac
{\rho\ell^{2}}{2\left(  d-2\right)  ^{2}}\kappa^{i}\kappa^{j}g_{ij}^{\left(
0\right)  }+...\right)  . \label{Extrinsic_K}%
\end{equation}
Finally, in order to evaluate $S_{EE}^{ct}$, we consider the expansion of
$\sqrt{\widetilde{\gamma}}\widetilde{\gamma}^{ab}k_{ab}$ at the cutoff
$\rho=\varepsilon$, where the limit of $\varepsilon\rightarrow0$ has to be
evaluated at the end. Thus, we have%

\begin{equation}
\left.  \sqrt{\widetilde{\gamma}}\widetilde{\gamma}^{ab}k_{ab}\right\vert
_{\rho=\varepsilon}=\frac{\left(  d-2\right)  \sqrt{\sigma^{\left(  0\right)
}}}{\ell\varepsilon^{\left(  d-2\right)  /2}}\left(  1+\varepsilon\left[
\frac{\left(  d-4\right)  }{2\left(  d-2\right)  }tr\left[  \sigma^{\left(
2\right)  }\right]  -\frac{\ell^{2}}{2\left(  d-2\right)  ^{2}}\kappa
^{i}\kappa^{j}g_{ij}^{\left(  0\right)  }\right]  +...\right)  .
\end{equation}

Now we have all the pieces required to check that the divergences of
$S_{EE}^{ren}$ vanish. We therefore consider that
\begin{equation}
S_{EE}^{RT}=\frac{1}{4G}\int\limits_{\Sigma}d^{2}y\sqrt{\gamma}=\frac{1}%
{4G}\int_{\partial\Sigma_{\varepsilon}}dy\int_{\varepsilon}^{\rho_{\text{max}%
}}d\rho\sqrt{\gamma},
\end{equation}
where $\rho_{\text{max}}$ is the maximum value of $\rho$ in the $\Sigma$
minimal surface, which depends on the choice of entangling surface at the
conformal boundary. By subsuming the finite part of the $\rho$ integral in a
constant $C$, we can therefore write that%

\begin{equation}
\int_{\varepsilon}^{\rho_{\text{max}}}d\rho\sqrt{\gamma}=C+\frac{\ell
\sqrt{\sigma^{\left(  0\right)  }}}{\left(  d-2\right)  \varepsilon^{\left(
d-2\right)  /2}}\left(  1+\varepsilon\left[  \frac{\left(  d-2\right)
}{2\left(  d-4\right)  }tr[\sigma^{\left(  2\right)  }]+\frac{\ell^{2}%
}{2\left(  d-4\right)  \left(  d-2\right)  }\kappa^{i}\kappa^{j}%
g_{ij}^{\left(  0\right)  }\right]  +...\right)  .
\end{equation}
And in our particular case, for 3D CFTs, we note that only the leading terms
in the $\varepsilon$ expansion contribute to the divergences. Thus, we have%

\begin{equation}
\left.  \sqrt{\widetilde{\gamma}}\widetilde{\gamma}^{ab}k_{ab}\right\vert
_{\rho=\varepsilon}=\frac{\sqrt{\sigma^{\left(  0\right)  }}}{\ell
\varepsilon^{1/2}}~;~\int_{\varepsilon}^{\rho_{\text{max}}}d\rho\sqrt{\gamma
}=C+\frac{\ell\sqrt{\sigma^{\left(  0\right)  }}}{\varepsilon^{1/2}},
\end{equation}
and therefore, the structure of divergences of $S_{EE}^{ren}$, in the limit of
$\varepsilon\rightarrow0$, give the following expression
\begin{equation}
S_{EE}^{ren}=\underset{0}{\lim_{\varepsilon\rightarrow0}\underbrace{\frac
{1}{4G}\int\limits_{\partial\Sigma_{\varepsilon}}dy\frac{\ell\sqrt
{\sigma^{\left(  0\right)  }}}{\varepsilon^{1/2}}-\frac{\ell^{2}}{4G}%
\int\limits_{\partial\Sigma_{\varepsilon}}dy\frac{\sqrt{\sigma^{\left(
0\right)  }}}{\ell\varepsilon^{1/2}}}}+\frac{C}{4G},
\end{equation}
where $\frac{C}{4G}$ is $O\left(  1\right)  $. We have therefore verified,
explicitly, that the divergences in $S_{EE}^{ren}$ cancel each other for any
entangling surface $\partial A$, and thus, $S_{EE}^{ren}$ is correctly renormalized.

Finally, we show that our expression for $S_{EE}^{ren}$ (exhibited in
eq.(\ref{S_EE^ren})) is equivalent to the expression obtained in \cite{Marika}
and presented in eq.(\ref{S_ren_usual_4D}). To see this, we consider that%

\begin{equation}
k_{b}^{a}=\widetilde{\gamma}^{ac}k_{cb}=\frac{1}{\ell}\left(  \delta_{b}%
^{a}-\rho\left[  \left(  \sigma^{\left(  2\right)  }\right)  _{b}^{a}%
+\frac{\ell^{2}}{2\left(  d-2\right)  ^{2}}\kappa^{i}\kappa^{j}g_{ij}^{\left(
0\right)  }\delta_{b}^{a}\right]  +...\right)  ,
\end{equation}
and therefore, for $d=3$, we have that
\begin{equation}%
\begin{tabular}
[c]{l}%
$\left.  B_{1}\right\vert _{\varepsilon}=-2dy\sqrt{\left.  \widetilde{\gamma
}\right\vert _{\varepsilon}}\delta_{a}^{b}k_{b}^{a}=-\frac{2}{\ell}%
dy\sqrt{\left.  \widetilde{\gamma}\right\vert _{\varepsilon}}\left(
\underset{1}{\underbrace{\delta_{a}^{a}}}+O\left(  \varepsilon\right)
\right)  ,$\\
$S_{EE}^{ren}=\frac{Vol\left(  \Sigma\right)  }{4G}-\frac{\ell}{4G}%
\int\limits_{\partial\Sigma}dy\sqrt{\widetilde{\gamma}},$%
\end{tabular}
\end{equation}
thus recovering the known result.

\subsection{Topological interpretation of the renormalized EE}

We now give a topological interpretation of $S_{EE}^{ren}$, considering the
Euler theorem given in eq.(\ref{Euler_Theo}), and the definition of the
curvature for the AdS group, which for an AAdS manifold is given by%

\begin{equation}
\left(  \mathcal{F}_{AdS}\right)  _{\nu_{1}\nu_{2}}^{\mu_{1}\mu_{2}}%
=R_{\nu_{1}\nu_{2}}^{\mu_{1}\mu_{2}}+\frac{1}{\ell^{2}}\delta_{\left[  \nu
_{1}\nu_{2}\right]  }^{\left[  \mu_{1}\mu_{2}\right]  },\label{AdS_curvature}%
\end{equation}
where $R_{\nu_{1}\nu_{2}}^{\mu_{1}\mu_{2}}$ is the Riemann tensor of the manifold.

Using eq.(\ref{Euler_Theo}), the Chern form that plays the role of counterterm
for EE can be expressed as%

\begin{equation}%
{\displaystyle\int\limits_{\partial\Sigma}}
B_{1}=%
{\displaystyle\int\limits_{\Sigma}}
d^{2}y\sqrt{\gamma}\mathcal{R}-\frac{\ell^{2}}{8G}\left(  4\pi\chi\left(
\Sigma\right)  \right)  ,
\end{equation}
where $\mathcal{R}$ is the Ricci scalar for the induced metric $\gamma_{ab}$
on $\Sigma$. Therefore, we can write $S_{EE}^{ren}$ as%

\begin{equation}
S_{EE}^{ren}=\frac{1}{4G}%
{\displaystyle\int\limits_{\Sigma}}
d^{2}y\sqrt{\gamma}+\frac{\ell^{2}}{8G}%
{\displaystyle\int\limits_{\Sigma}}
d^{2}y\sqrt{\gamma}\mathcal{R}-\frac{\ell^{2}}{8G}\left(  4\pi\chi\left(
\Sigma\right)  \right)  .
\end{equation}
The above formula can be rewritten as%

\begin{equation}
S_{EE}^{ren}=\frac{\ell^{2}}{8G}%
{\displaystyle\int\limits_{\Sigma}}
d^{2}y\sqrt{\gamma}\left(  \mathcal{R}+\frac{2}{\ell^{2}}\right)  -\frac
{\pi\ell^{2}}{2G}\chi\left(  \Sigma\right)  ,
\end{equation}
and using the properties of the totally antisymmetric Kronecker delta, we obtain%

\begin{equation}
S_{EE}^{ren}=\frac{\ell^{2}}{16G}%
{\displaystyle\int\limits_{\Sigma}}
d^{2}y\sqrt{\gamma}\delta_{\left[  a_{1}a_{2}\right]  }^{\left[  b_{1}%
b_{2}\right]  }\underset{=\left(  \left.  \mathcal{F}_{AdS}\right\vert
_{\Sigma}\right)  _{b_{1}b_{2}}^{a_{1}a_{2}}}{\underbrace{\left(
\mathcal{R}_{b_{1}b_{2}}^{a_{1}a_{2}}+\frac{1}{\ell^{2}}\delta_{\left[
b_{1}b_{2}\right]  }^{\left[  a_{1}a_{2}\right]  }\right)  }}-\frac{\pi
\ell^{2}}{2G}\chi\left(  \Sigma\right)  .\label{S_EE^Topol}%
\end{equation}

The expression given in eq.(\ref{S_EE^Topol}) is interesting because it makes
manifest the connections of the renormalized EE with the topology of the
extremal surface $\Sigma$, and also to its algebraic-geometrical properties as
an AAdS Riemannian submanifold. In particular, we can recognize the curvature
of the AdS group \cite{OleaF} for $\Sigma$, denoted $\left(  \left.
\mathcal{F}_{AdS}\right\vert _{\Sigma}\right)  _{b_{1}b_{2}}^{a_{1}a_{2}}$,
and also its Euler characteristic $\chi\left(  \Sigma\right)  $.

\subsection{Explicit example: Disk-like entangling region in CFT$_{3}$, with a
global AdS$_{4}$ bulk\label{ExplicitExample}}

We now compute $S_{EE}^{ren}$ for the particular case of a disc-like
entangling region in the ground state of a 3D CFT, which is dual to a global
AdS$_{4}$ bulk on the gravity side, using our topological interpretation of
the renormalized EE given in eq.(\ref{S_EE^Topol}). The importance of this
example is explained in detail in section \ref{Outlook}, but here we only
mention that $S_{EE}^{ren}$ is related to the $F$ quantity \cite{F-theo} by
$S_{EE}^{ren}=-F$ and to the $a$-charge \cite{S-c-Myers} by $S_{EE}%
^{ren}=-2\pi a_{3}$. Both of these order parameters of the CFT that are
conjectured to decrease along RG flows between conformal fixed points (and can
be thought of as generalizations of Zamolodchikov's c-theorem \cite{cTheo}).

We start by considering the global AdS$_{4}$ bulk metric, which can be written
in polar coordinates as%

\begin{equation}
ds_{G}^{2}=\frac{\ell^{2}d\rho^{2}}{4\rho^{2}}+\frac{1}{\rho}\left(
-dt^{2}+dr^{2}+r^{2}d\phi^{2}\right)  =G_{\mu\nu}dx^{\mu}dx^{\nu}.
\end{equation}
Then, as it is shown in the appendix \ref{Verification}, the minimal surface
in the bulk, which has a boundary that is conformal to the circle which
constitutes the entangling surface, is given by the parametrization:
\begin{equation}
\Sigma:\left\{  t=const~;~r^{2}+\ell^{2}\rho=R^{2}\right\}  ,
\end{equation}
where $R$ is the radius of the circle. Now, we compute the induced metric on
$\Sigma$, defined in eq.(\ref{InducedMetricSigma}), considering that the
coordinates on $\Sigma$ are $y^{a}=\left\{  \rho,\phi\right\}  $, and those in
the bulk are given by $x^{\mu}=\left\{  \rho,t,r,\phi\right\}  $. Then, for
the induced metric on $\Sigma$, we obtain%

\begin{equation}
ds_{\gamma}^{2}=\frac{\ell^{2}}{4\rho^{2}}\left(  1+\frac{\ell^{2}\rho
}{\left(  R^{2}-\ell^{2}\rho\right)  }\right)  d\rho^{2}+\frac{\left(
R^{2}-\ell^{2}\rho\right)  }{\rho}d\phi^{2}=\gamma_{\alpha\beta}dy^{\alpha
}dy^{\beta}.
\end{equation}

Given the induced metric on $\Sigma$, we compute its AdS curvature $\left(
\left.  \mathcal{F}_{AdS}\right\vert _{\Sigma}\right)  _{b_{1}b_{2}}%
^{a_{1}a_{2}}$ according to eq.(\ref{AdS_curvature}), and we find that it
vanishes identically. Also, we note that $\Sigma$ is topologically equivalent
to a disk, and therefore, $\chi\left(  \Sigma\right)  =1$. Thus, using the
topological expression for $S_{EE}^{ren}$ given in eq.(\ref{S_EE^Topol}), we obtain%

\begin{equation}
S_{EE}^{ren}=-\frac{\pi\ell^{2}}{2G},
\end{equation}
in agreement with the result obtained in \cite{Marika}.

Therefore, as further explained in section \ref{Outlook}, we have that for the
3D CFT in the ground state, in terms of the quantities on the gravity side,
$F=\frac{\ell^{2}\pi}{2G_{4}}$ and $a_{3}=\frac{\ell^{2}}{4G_{4}}$, in
agreement with the previously known results. We mention however that in our
computation we were able to exhibit new properties of the EE that, to the best
of our knowledge, had not been noticed before.

\section{Outlook\label{Outlook}}

So far, we have presented a new prescription for computing $S_{EE}^{ren}$ in
eq.(\ref{S_EE^ren}), which was derived directly from the replica trick by
considering a suitably renormalized bulk gravity action (eq.(\ref{I^ren_4D})).
We have also verified the finiteness of the EE obtained through such
prescription, and its equivalence with the known result given in
\cite{Marika}. Furthermore, in eq.(\ref{S_EE^Topol}), we have interpreted the
result for $S_{EE}^{ren}$ in terms of the topological properties of the
minimal surface $\Sigma$, and its geometrical properties as an AAdS submanifold.

EE, as considered in Quantum Information Theory for systems with finite
dimensional Hilbert Spaces, is positive definite and can be computed directly
as shown in eq.(\ref{VonNewmann}). As it was mentioned in section
\ref{EE_Replica}, it encodes the level of entanglement between a quantum
subsystem $A$ and its complement ($A^{c}$). In the case of CFTs (and more
generally QFTs), the infinite-dimensional Hilbert space of the theory
introduces the usual UV divergence in the EE. In the gravity side, this
divergence appears in the area of the minimal surface $\Sigma$ due to the
infinite conformal factor in the metric at the spacetime boundary. As it is
explicitly shown in section \ref{ExplicitExample}, the renormalized EE
($S_{EE}^{ren}$) is no longer positive definite, so its physical
interpretation as an order parameter and its interest for the study of CFTs
needs to be explicitated.

In particular, as mentioned in \cite{Marika}, the renormalized entanglement
entropy $S_{EE}^{ren}$ is of interest because of its connection with
quantities that are important for the study of holographic renormalization
group (RG) flows. For example, in the case of a 3D CFT at the boundary and a
disc-shaped entangling region, $S_{EE}^{ren}=-F$, where the F quantity is
defined in terms of the renormalized partition function of the theory on a
three sphere as $F=-\ln Z_{S^{3}}$, and it decreases along RG flows
\cite{F-theo}. Also, the renormalized EE for 3D CFTs with a disc-shaped
entangling region can be written in terms of the $a$-charge of the CFT as
$S_{EE}^{ren}=-2\pi a_{3}$, where $a_{3}$ is conjectured to satisfy the
relation that $\left(  a_{3}\right)  _{UV}\geq\left(  a_{3}\right)  _{IR}$,
for any RG flow between conformal fixed points, as discussed in
\cite{S-c-Myers}. Therefore with our method, we recover the known results
\cite{Marika} of $F=\frac{\ell^{2}\pi}{2G_{4}}$ and $a_{3}=\frac{\ell^{2}%
}{4G_{4}}$, which can be translated in terms of the CFT quantities using the
standard holographic dictionary ($G_{4}$ is the gravitational constant of the
4D AAdS bulk). All these quantities that (are conjectured to) decrease along
RG flows can be considered as generalizations of Zamolodchikov's c-theorem
\cite{cTheo}. They encode information about the number of degrees of freedom,
which decreases as the theory flows to the infrared (IR).

Another quantity that is related to the EE and is useful for characterizing
the informational content of CFTs is the Mutual Information (MI)
\cite{Mutual1}\cite{Mutual2}, which is defined in terms of differences of EEs as%

\begin{equation}
I_{A,B}=S_{EE}\left(  A\right)  +S_{EE}\left(  B\right)  -S_{EE}\left(  A\cup
B\right)  , \label{Mutual}%
\end{equation}
where $I_{A,B}$ denotes the MI between regions A and B, and $S_{EE}\left(
X\right)  $ denotes the EE of the entangling region X. If one instead
considers $S_{EE}^{ren}\left(  X\right)  $ as the EE for region X, the result
of $I_{A,B}$ is left unchanged for regions that do not overlap. Therefore, the
renormalized EE can be used in the computation of MI without changing its
properties. In particular, even if $S_{EE}^{ren}$ can be negative, $I_{A,B}$
is always positive definite. This is important because it is usually the MI
that is used when characterizing the amount of correlation between different
regions in a CFT. For example, the MI can be used to place bounds on
correlators of operators defined on separate regions \cite{Mutual2}.

As it will be described in a follow-up paper, we can extend the method for
computing $S_{EE}^{ren}$ to AAdS manifolds of arbitrary even dimensions by
considering the renormalized Euclidean action given in \cite{K1Even}, and by
repeating the replica procedure. As future work, we will also study how to
extend the scheme to AAdS manifolds of arbitrary odd dimensions, considering
the renormalized Euclidean gravitational action discussed in \cite{K2Odd}; and
also to higher-curvature theories of gravity, specially those of the Lovelock
class \cite{Lovelock1}\cite{Lovelock2}.

We will also study the application of our renormalization procedure to other
QIT measures, like the Entanglement Renyi Entropies (EREs) \cite{RenyiXiDong}%
-\cite{MyersRenyi} and the complexity \cite{AliComplexity}-\cite{IgnacioReyes}%
, for CFTs of arbitrary dimensions with AAdS gravity duals. Regarding the
complexity, we mention that an example of topological renormalization has
already been achieved in \cite{IgnacioReyes}, although only for the particular
case of $AdS_{3}/CFT_{2}$.

\begin{acknowledgments}
The authors thank Y. Novoa for interesting discussions. G.A. is a Universidad
Andres Bello (UNAB) Ph.D. Scholarship holder, and his work is supported by
Direcci\'{o}n General de Investigaci\'{o}n (DGI-UNAB). This work is funded in
part by FONDECYT Grant No. 1170765, UNAB Grant DI-1336-16/R and CONICYT Grant
DPI 20140115.
\end{acknowledgments}

\appendix

\section{Derivation of the minimal area condition in global AdS}

In section \ref{ExplicitExample}, we consider that the minimal surface
$\Sigma$ in the global AdS$_{4}$ bulk for a disk-like entangling region in the
dual 3D CFT in its ground state is given by $\Sigma:\left\{  t=const~;~r^{2}%
+\ell^{2}\rho=R^{2}\right\}  $, where $R$ is the radius of the disc. Here, we
proceed to explicitly justify this claim. We first derive the minimal surface
condition, in the form of an Euler-Lagrange differential equation that has to
be obeyed by the embedding function of the minimal surface $\Sigma$, and then
we proceed to check that the surface $\Sigma$ as defined above does indeed
satisfy this condition. We note that this analysis is standard, and the reason
why we repeat it here is because there is a small mistake in the treatment
done by Taylor and Woodhead, presented in eq.(3.12) of \cite{Marika}.

We first consider the metric of global AdS$_{D}$, written in cartesian coordinates:%

\begin{equation}
ds_{G}^{2}=G_{\mu\nu}dx^{\mu}dx^{\nu}=\frac{\ell^{2}d\rho^{2}}{4\rho^{2}%
}+\frac{-dt^{2}+\delta_{ab}dx^{a}dx^{b}}{\rho}.
\end{equation}
Then, we consider the parametrization of a codimension-2 surface $\Sigma$,
with worldvolume coordinates given by $\tau$ and $y^{a}$ at $t=const$. The
embedding is done in the static gauge, such that $\rho=\tau$ and $x^{a}=y^{a}%
$, for $a=1,...,D-3$, and$~x^{D-2}=z\left(  \rho,x^{a}\right)  $, where $z$ is
the embedding function and $D$ is the dimension of the bulk manifold. Then,
the induced metric $\gamma_{ab}$ is given by $\gamma_{ab}=\frac{\partial
x^{\mu}}{\partial y^{a}}\frac{\partial x^{\nu}}{\partial y^{b}}G_{\mu\nu}$,
and in terms of the embedding function $z\left(  \rho,x^{a}\right)  $, we obtain%

\begin{equation}%
\begin{tabular}
[c]{l}%
$\gamma_{\rho\rho}=G_{\rho\rho}+z_{,\rho}z_{,\rho}G_{zz}=\frac{\ell^{2}}%
{4\rho^{2}}+\frac{z_{,\rho}z_{,\rho}}{\rho},$\\
$\gamma_{ab}=G_{ab}+z_{,a}z_{,b}G_{zz}=\frac{1}{\rho}\left(  \delta
_{ab}+z_{,a}z_{,b}\right)  ,$\\
$\gamma_{\rho a}=z_{,\rho}z_{,a}G_{zz}=\frac{z_{,\rho}z_{,a}}{\rho}.$%
\end{tabular}
\ \label{Sigma_Metric_min}%
\end{equation}

Now, we can derive the minimal area condition. In order to do this, we
consider that $Vol\left(  \Sigma\right)  =%
{\displaystyle\int\limits_{\Sigma}}
d^{D-2}y\sqrt{\gamma}$ is the area functional, and we define the auxiliary
function $m\left(  \rho,x^{a}\right)  =\sqrt{1+4\frac{\rho}{\ell^{2}}z_{,\rho
}z_{,\rho}+z_{,a}z_{,a}}$, such that $Vol\left(  \Sigma\right)  =%
{\displaystyle\int\limits_{\Sigma}}
d^{D-2}y\frac{\ell m\left(  \rho,x^{a}\right)  }{2\rho^{\left(  D-1\right)
/2}}$. Then, we impose that the variation of the area functional with respect
to the embedding function $z\left(  \rho,x^{a}\right)  $ has to be zero, in
order for the surface $\Sigma$ to have extremal area. Therefore, the $z\left(
\rho,x^{a}\right)  $ corresponding to said minimal surface has to fulfill the
differential equation resulting from the extremization condition. To derive
the extremization condition, we consider that under the variation,%

\begin{equation}
\delta_{z}Vol\left(  \Sigma\right)  =%
{\displaystyle\int\limits_{\Sigma}}
d^{D-2}y\frac{\ell}{4m\left(  \rho,x^{a}\right)  \rho^{\left(  D-1\right)
/2}}\left(  8\frac{\rho}{\ell^{2}}z_{,\rho}\delta z_{,\rho}+2z_{,a}\delta
z_{,a}\right)  ,
\end{equation}
and then, requiring that $\delta_{z}Vol\left(  \Sigma\right)  =0$ and
integrating by parts, we obtain the corresponding Euler-Lagrange condition
given by%

\begin{equation}
\partial_{a}\left(  \frac{z_{,a}}{4\rho^{\left(  D-1\right)  /2}m\left(
\rho,x^{a}\right)  }\right)  +\partial_{\rho}\left(  \frac{z_{,\rho}}{\ell
^{2}\rho^{\left(  D-3\right)  /2}m\left(  \rho,x^{a}\right)  }\right)  =0.
\label{Extremal_Area_Condition}%
\end{equation}
Thus, in order for $\Sigma$ to be the minimal surface, its embedding function
$z\left(  \rho,x^{a}\right)  $ has to satisfy
eq.(\ref{Extremal_Area_Condition}). To argue that $\Sigma$ is a minimum, and
not a maximum, we note that due to the divergent conformal factor in the
metric at $\rho\rightarrow0$, the maximum is not well defined (intuitively, it
would be a surface located entirely at the boundary of spacetime). Of course,
there may be more than one surface $\Sigma$ that satisfies
eq.(\ref{Extremal_Area_Condition}), which would mean that these are multiple
local minima of the area functional. In such a case, the true minimal surface
is the one among them that has the smallest value for the area (after a
suitable renormalization, by, for example, the method described in the body of
this paper).

Now, in the next section, we verify that the surface $\Sigma$ of section
\ref{ExplicitExample} does indeed satisfy the extremal area condition of
eq.(\ref{Extremal_Area_Condition}).

\section{Verification that the $\Sigma$ considered for a disk-like entangling
region is the minimal surface\label{Verification}}

In the case of global AdS$_{4}$, we consider a surface $\Sigma$ parametrized
as $\Sigma:\left\{  t=const~;~r^{2}+\ell^{2}\rho=R^{2}\right\}  $, and we
proceed to show that it satisfies the extremal area condition of
eq.(\ref{Extremal_Area_Condition}). We first write the corresponding embedding
function $z\left(  \rho,x\right)  $ in cartesian coordinates, considering that
$r^{2}=x^{2}+z^{2}$. Thus we have that $z\left(  \rho,x\right)  =\pm
\sqrt{R^{2}-\ell^{2}\rho-x^{2}}$. Then, computing the derivatives of the
embedding function, we have that%

\begin{equation}%
\begin{tabular}
[c]{l}%
$z_{,a}=\mp\frac{x}{\sqrt{R^{2}-\ell^{2}\rho-x^{2}}}~;~z_{,\rho}=\mp\frac
{\ell^{2}}{2\sqrt{R^{2}-\ell^{2}\rho-x^{2}}},$\\
$m\left(  \rho,x\right)  =\sqrt{1+4\frac{\rho}{\ell^{2}}z_{,\rho}z_{,\rho
}+z_{,a}z_{,a}}=\frac{R}{\sqrt{R^{2}-\ell^{2}\rho-x^{2}}},$%
\end{tabular}
\end{equation}
and replacing the corresponding terms into eq.(\ref{Extremal_Area_Condition}),
we have that%

\begin{equation}
\partial_{a}\left(  \frac{z_{,a}}{4\rho^{\left(  D-1\right)  /2}m\left(
\rho,x^{a}\right)  }\right)  +\partial_{\rho}\left(  \frac{z_{,\rho}}{\ell
^{2}\rho^{\left(  D-3\right)  /2}m\left(  \rho,x^{a}\right)  }\right)
=\pm\left(  -\frac{1}{2R}\partial_{\rho}\left(  \frac{1}{\rho^{1/2}}\right)
-\frac{1}{4R\rho^{3/2}}\right)  =0,
\end{equation}
and therefore, $\Sigma$ is indeed the minimal surface.

\end{document}